\documentclass[11pt]{article}
\usepackage[cp1250]{inputenc}
\usepackage{amsmath}
\usepackage{amsfonts}
\usepackage{graphicx}
\usepackage{yfonts}
\usepackage{wasysym}
\usepackage{color}
\usepackage[normalem]{ulem}
\usepackage{amsthm}
\usepackage{bm}
\usepackage{bbm}
\usepackage{mathtools}

\DeclareMathOperator{\Tr}{Tr}
\DeclareMathOperator{\imag}{i}

\DeclareMathOperator{\der}{d}
\newcommand\bigforall{\mbox{\Large $\mathsurround=1pt\forall$}}

\def\<{\langle}
\def\>{\rangle}
\newtheorem{thm}{Theorem}
\newtheorem{define}{Definition}

\newtheorem{example}{Example}
\usepackage{placeins}
\def\oper{{\mathchoice{\rm 1\mskip-4mu l}{\rm 1\mskip-4mu l}
{\rm 1\mskip-4.5mu l}{\rm 1\mskip-5mu l}}}

\title{\bf Two definitions of the Gell-Mann channels\\ -- a comparative analysis}
\author{ Katarzyna Siudzi{\'n}ska\\ Institute of Physics, Nicolaus Copernicus University, \\ ul. Grudzi\c{a}dzka 5, 87-100 Toru{\'n}, Poland }
\date{\today}

\begin{document}

\maketitle

\begin{abstract}
Our goal is to compare two unequivalent definitions of the Gell-Mann channels. It turns out that both definitions coincide for qubits and qutrits. In higher dimensions, there exist some constraints under which the channels describe the same dynamics. Finally, we find the GKSL time-local generators for a class of the Gell-Mann channels.
\end{abstract}

\section{Introduction}

There are two methods of describing the time-evolution of open quantum systems: in the language of quantum channels and quantum master equations. Quantum channels $\Lambda$ are completely positive, trace-preserving maps, where we know that a map is completely positive if and only if it can be written in the Kraus form \cite{Kraus},
\begin{equation}\label{Kraus_form}
\Lambda [X]=\sum_\alpha K_\alpha^\dagger X K_\alpha.
\end{equation}
The time-dependent quantum channel, also known as the dynamical map, is the solution of the master equation
\begin{equation}\label{master_equation}
\frac{\der}{\der t}\Lambda(t)=\mathcal{L}\Lambda(t)
\end{equation}
with the condition $\Lambda(0)=\oper$, where $\mathcal{L}$ is the (usually time-dependent) generator of evolution. If $\mathcal{L}$ does not depend on time, then it is the generator of the quantum dynamical semigroup and can be written in the GKSL form \cite{GKS, L},
\begin{equation}\label{GKLS}
\mathcal{L}[X]=\sum_\alpha\left(V_\alpha^\dagger XV_\alpha-\frac 12 \{V_\alpha V_\alpha^\dagger,X\}\right),
\end{equation}
where we skipped the Hamiltonian part $-\imag[H,X]$.

As an example of the evolution that has been well-analyzed in both approaches, one can consider the dynamics of a qubit. The corresponding quantum channel is defined by
\begin{equation}\label{Pauli_channel}
\Lambda_P(t)[\sigma_\alpha]=\lambda_\alpha\sigma_\alpha,
\end{equation}
where $\sigma_\alpha$ are the Pauli matrices. It is well-known that (\ref{Pauli_channel}) is completely positive if and only if its eigenvalues satisfy the Fujiwara-Algoet condition \cite{Fujiwara},
\begin{equation}\label{Fuji}
|\lambda_0\pm\lambda_3|\geq|\lambda_1\pm\lambda_2|,
\end{equation}
and (\ref{Pauli_channel}) is trace-preserving for $\lambda_0=1$. The Pauli channel is often defined by its Kraus form,
\begin{equation}\label{Pauli_Kraus}
\Lambda_P(t)[X]=\sum_\alpha p_\alpha(t)\sigma_\alpha X \sigma_\alpha,
\end{equation}
with $p_\alpha(t)\geq 0$ and $\sum_\alpha p_\alpha(t)=1$. The time-local generator of the Pauli channel can be written in the GKSL form as
\begin{equation}\label{Pauli_generator}
\mathcal{L}_P(t)[X]=\sum_\alpha\gamma_\alpha(t)\left(\sigma_\alpha X\sigma_\alpha-X\right).
\end{equation}
There are many generalizations of the Pauli channel to higher dimensions, like the generalized Pauli channels or the Weyl channels (for recent reviews see \cite{Ruskai, Ohno, Publ3}, and also \cite{Filip, Filip2}).

In the present paper, we are analysing the Hermitian generalization of the Pauli channel with the (generalized) Gell-Mann matrices. Let us recall that the Gell-Mann matrices $\sigma_{ij}$ are given by
\begin{align}\label{Gell-Manns}
\bigforall_{0\leq i<j\leq n-1}\qquad&\sigma_{ij}:=e_{ij}+e_{ji},\\
\bigforall_{0\leq i<j\leq n-1}\qquad&\sigma_{ji}:=-\imag(e_{ij}-e_{ji}),\\
\bigforall_{0\leq j\leq n-1} \qquad &\sigma_{jj}:=\sqrt{\frac{2}{j(j+1)}}\left(\sum_{i<j}e_{ii}-je_{jj}\right),\\
&\sigma_{00}:=\sum_{j=0}^{n-1}e_{jj}.
\end{align}
In the next section, we are going to show that the higher-dimensional analogues of definitions (\ref{Pauli_channel}) and (\ref{Pauli_Kraus}) are equivalent only for $n\leq 3$. Our main goal is to analyze and compare the two definitions and find the Fujiwara-Algoet conditions for $n\geq 2$. For a subclass of channels, we would also like to check when the time-local generator of evolution has the GKSL form. Where time-dependence of coefficients is not explicitly stated, it it assumed that they all depend on the same $t\geq 0$. The proofs to all the theorems are presented in the appendices.

\section{Two channels}

There are two natural ways in which one could introduce the Hermitian generalization of the Pauli channel (\ref{Pauli_channel}). The first idea that comes to mind is to use the Kraus representation. Let us introduce the mapping
$\Lambda_{KF}(t):\mathcal{B}(\mathcal{H}_n)\to\mathcal{B}(\mathcal{H}_n)$, where the Kraus operators are given by the (generalized) Gell-Mann matrices,

\begin{equation}\label{lambda_KF}
\Lambda_{KF}(t)[X]:=\sum_{i,j=0}^{n-1}p_{ij}(t)\sigma_{ij}X\sigma_{ij}.
\end{equation}
$\Lambda_{KF}(t)$ is a quantum channel if it is a completely positive, trace-preserving (CPT) map. Complete positivity is equivalent to the requirement that the time-dependent coefficients are non-negative, i.e. $p_{ij}(t)\geq 0$, and being trace-preserving imposes the additional condition $\Tr\Lambda_{KF}[X]=\Tr X$.

The set $\{\Lambda_{KF}(t)\,|\,t\geq 0\}$ is a one-parameter family of dynamical maps if for every $t\geq 0$, $\Lambda_{KF}(t)$ is a quantum channel and, additionally, $\Lambda_{KF}(0)=\oper$. By $\oper$, we understand the identity mapping, $\oper [X]=X$. The last condition is guaranteed by taking $p_{00}(0)=1$ as the only non-vanishing coefficient at $t=0$.

Although it may seem counterintuitive, the condition for the map $\Lambda_{KF}(t)$ to be trace-preserving is, in fact, a set of equations for the time-dependent coefficients. This matter is discussed in more details in Theorem \ref{TP_theorem}.

\begin{thm}\label{TP_theorem}
The map $\Lambda_{KF}(t)$ is trace-preserving if and only if the time-dependent coefficients $p_{ij}(t)$ satisfy the following set of $n$ equations:
\begin{align}
&p_{00} =1-\sum_{0<l<n}\tilde{p}_{0l}-2\sum_{0<l<n}\frac{1}{l(l+1)}p_{ll},\label{TP_conditions1}\\
&\sum_{1<l<n}(\tilde{p}_{1l}-\tilde{p}_{0l})=0,\label{TP_conditions2}\\
&p_{22}=p_{11}+\tilde{p}_{01}-\tilde{p}_{12}+\sum_{2<l<n}(\tilde{p}_{0l}-\tilde{p}_{2l})\label{TP_conditions3},\\
&p_{kk}=p_{22}+\sum_{1<j<k}\frac{j+1}{2j}\left[
\sum_{0\leq l<j}(\tilde{p}_{lj}-\tilde{p}_{l,j+1})
+\sum_{j+1<l<n}(\tilde{p}_{jl}-\tilde{p}_{j+1,l})
\right]\quad\bigforall_{2<k<n}\label{TP_conditions4},
\end{align}
where $\tilde{p}_{ij}:={p}_{ij}+{p}_{ji}$ for all $i<j$.
\end{thm}

Finally, we arrive at the following definition.

\begin{define}\label{def_GM_Kraus}
The mapping $\Lambda_{KF}(t)$ given by the Kraus form (\ref{lambda_KF}), with non-negative coefficients $p_{ij}(t)$ satisfying the requirements (\ref{TP_conditions1}-\ref{TP_conditions3}) together with $p_{00}(0)=1$ and $p_{ij}(0)=0$, defines the dynamical map called the KF Gell-Mann channel.
\end{define}

Although nothing is wrong with Definition \ref{def_GM_Kraus}, there is a major drawback in defining a channel by its Kraus representation. While it is easy to determine whether the map in the Kraus form is CPT, there is no effective procedure of finding the Kraus operators. Moreover, the Kraus representation for a given channel is not unique.

Now is the moment to talk about the other natural Hermitian generalization -- with the eigenvalue equations. Assume that the Gell-Mann matrices are the eigenvectors of the mapping $\Lambda_{EV}(t):\mathcal{B}(\mathcal{H}_n)\to\mathcal{B}(\mathcal{H}_n)$. Then, this map is fully and uniquely determined by the following equations:
\begin{equation}\label{lambda_EV}
\Lambda_{EV}(t)[\sigma_{\alpha\beta}]=\lambda_{\alpha\beta}(t) \sigma_{\alpha\beta},
\end{equation}
where $\lambda_{ij}(t)$ are the eigenvalues of $\Lambda_{EV}(t)$. To find the necessary and sufficient requirements for $\Lambda_{EV}(t)$ to describe a completely positive map, we use the method presented in \cite{Lee}. It allows us to formulate the following theorem.

\begin{thm}\label{CP_theorem}
The map $\Lambda_{EV}(t)$ is CP if and only if its eigenvalues $\lambda_{ij}(t)$ satisfy these \mbox{$n(n-1)/2+1$} conditions:
\begin{align}
&|\lambda_{ij}-\lambda_{ji}|\leq\left|\lambda_{00}^D-\frac{2}{j+1}\lambda_{jj}
+\sum_{j<k<n}\frac{2}{k(k+1)}\lambda_{kk}\right|,\label{CP_condition_1}\\
&\det A\geq 0,\label{CP_condition_2}
\end{align}
where the matrix elements $A_{ij}$ of $A$ are defined by
\begin{align}
&A_{ij}:=\lambda_{ij}+\lambda_{ji},\label{A_ij}\\
&A_{jj}:=\lambda_{00}+\frac{j}{j+1}\lambda_{jj}+\sum_{j<k<n}\frac{2}{k(k+1)}\lambda_{kk}\label{A_jj}.
\end{align}
\end{thm}

Being trace-preserving is much easier to verify, as the Gell-Mann matrices -- minus the identity -- are traceless. Therefore, we see that $\Lambda_{EV}(t)$ is TP if and only if $\lambda_{00}(t)=1$, and it is a dynamical map if, in addition, $\lambda_{ij}(0)=1$ for all $i$, $j$.

Now, let us introduce the proper definition.

\begin{define}\label{def_GM_values}
The mapping $\Lambda_{EV}$ given by the eigenvalue equations (\ref{lambda_EV}) and whose eigenvalues $\lambda_{ij}(t)$ satisfy the requirements (\ref{CP_condition_1}, \ref{CP_condition_2}), together with $\lambda_{00}(t)=1$ and $\lambda_{ij}(0)=1$, defines the dynamical map called the EV Gell-Mann channel.
\end{define}

Definitions \ref{def_GM_Kraus} and \ref{def_GM_values} are not equivalent. The question that poses itself is, which one should be treated as the proper definition of the Gell-Mann channel? Do we have any references to the Gell-Mann channels in the literature that would direct us in our choice? 

The answer to the second question is in the affirmative. There are, indeed, two recent articles by Hu and Fan \cite{HuFan1, HuFan2} dealing with 3-dimensional Gell-Mann channels. They are defined with the use of the Kraus operators, and then it is assumed that their eigenvectors are the Gell-Mann matrices. In other words, the authors use the channels that satisfy both our definitions. It is worth noting that for $n=2$, $3$, Definitions \ref{def_GM_Kraus} and \ref{def_GM_values} are equivalent, as we will see in Theorems \ref{p2lambda} and \ref{lambda2p}. Therefore, from now on, we treat $\Lambda_{KF}(t)$ and $\Lambda_{EV}(t)$ as two types of the Gell-Mann channels.

One could formulate the following problem: under what circumstances do $\Lambda_{KF}(t)$ and $\Lambda_{EV}(t)$ describe the same dynamics? The answer to this question is given in Theorems \ref{p2lambda} and \ref{lambda2p}.

\begin{thm}\label{p2lambda}
The KF Gell-Mann channel is the EV Gell-Mann channel if and only if its coefficients $p_{ij}(t)$ fulfil the following $(n-1)(n-2)/2$ conditions:
\begin{align}\label{p_cond}
\bigforall_{0\leq j,k<l\leq n-1}\quad
\tilde{p}_{jl}=\tilde{p}_{kl}=:\tilde{p}_l,
\end{align}
where $\tilde{p}_{kl}:=p_{kl} +p_{lk}$ for $k<l$. One can find the eigenvalues $\lambda_{ij}(t)$ by using the formulas below,
\begin{align}
&\lambda_{kl}=p_{kl}-p_{lk}+1-2p_{11} -\tilde{p}_1
-\sum_{j\neq l}\tilde{p}_j,\label{lambda_SA}\\
&\lambda_{kk}=1-(k+1) \tilde{p}_k-\sum_{k<j<n}\tilde{p}_j\label{lambda_D}.
\end{align}
\end{thm}

\begin{thm}\label{lambda2p}
The EV Gell-Mann channel is the KF Gell-Mann channel if and only if its eigenvalues $\lambda_{ij}(t)$ meet the following $n(n-1)/2$ requirements:
\begin{align}
&\bigforall_{0\leq j,k< l\leq n-1}\quad
\tilde{\lambda}_{jl}=\tilde{\lambda}_{kl}=:\tilde{\lambda}_l,\label{lambda_cond_1}\\
&\bigforall_{1\leq k\leq n-1}\quad
\lambda_{kk} =\lambda_{11} +\frac 12 \left(\tilde{\lambda}_1+\sum_{0<j<k}\tilde{\lambda}_j-k\tilde{\lambda}_k\right)\label{lambda_cond_2},
\end{align}
where $\tilde{\lambda}_{kl}:=\lambda_{kl}+\lambda_{lk}$. The coefficients $p_{ij}(t)$ are given by
\begin{align}
&p_{kl}=\frac 12 \left[\frac 1n -\frac{1}{l+1}\lambda_{ll}
+\sum_{l<j<n}\frac{1}{j(j+1)}\lambda_{jj} \right]
-\frac 14 \left[\lambda_{kl}-\lambda_{lk}\right],\label{p_SA}\\
&p_{kk} =\frac {1}{2n}\left[1-\lambda_{11} +n\lambda_{kk}
-\frac 12 \left(\tilde{\lambda}_1+\sum_{0<j<n}\tilde{\lambda}_j\right)\right],\label{p_D}\\
&p_{00} =\frac {1}{n^2}\left[1+\frac{n-1}{2}\left(2\lambda_{11}
+\tilde{\lambda}_1+\sum_{0<j<n}\tilde{\lambda}_j\right)\right]\label{p_0}.
\end{align}
\end{thm}

At this point, we think it beneficial for future understanding of the topic to present a simple example of the Gell-Mann channel that is KF but not EV. The lowest dimension where this is possible is $n=4$.

\begin{example}
Recall that the KF map is a quantum channel if and only if
\begin{align}\label{ex_1}
&\tilde{p}_{12}-\tilde{p}_{02}=\tilde{p}_{13}-\tilde{p}_{03},\\
&p_{22}=p_{11} +\tilde{p}_{01}-\tilde{p}_{12}+\tilde{p}_{03}-\tilde{p}_{23},\\
&p_{33}=p_{11} +\tilde{p}_{01}-\tilde{p}_{23}+\frac 12 (3\tilde{p}_{02}-2\tilde{p}_{12}-\tilde{p}_{03}),\\
&p_{00}=1+\frac 12 (\tilde{p}_{12}+\tilde{p}_{23})-\frac 32 (p_{11} +\tilde{p}_{01})-\frac 54 (\tilde{p}_{02}+\tilde{p}_{03}),
\end{align}
and it is, in addition, the EV channel as long as
\begin{equation}\label{ex_2}
\tilde{p}_{02}=\tilde{p}_{12}, \qquad \tilde{p}_{03}=\tilde{p}_{13}=\tilde{p}_{23}.
\end{equation}
Let us take the channel $\Lambda_{KF}(t)$ with
\begin{equation}\label{ex_3}
\tilde{p}_{02}=\tilde{p}_{12},\qquad \tilde{p}_{03}=\tilde{p}_{13}\neq \tilde{p}_{23}.
\end{equation}
Clearly, (\ref{ex_3}) satisfy (\ref{ex_1}) but not (\ref{ex_2}). Now, we check how such $\Lambda_{KF}(t)$ acts on the Gell-Mann matrices:
\begin{align*}
&\Lambda_{KF}(t) [\sigma_{00}]=\sigma_{00},\\
&\Lambda_{KF}(t) [\sigma_{11}]=(1-2\tilde{p}_{01}-\tilde{p}_{02}-\tilde{p}_{03})\sigma_{11},\\
&\Lambda_{KF}(t) [\sigma_{22}]=(1-3\tilde{p}_{02}-\tilde{p}_{03})\sigma_{22} +\frac{2\sqrt{3}}{3}
(\tilde{p}_{03}-\tilde{p}_{23})(e_{33}-e_{22}),\\
&\Lambda_{KF}(t) [\sigma_{33}]=(1-4\tilde{p}_{03})\sigma_{33} -\frac{2\sqrt{6}}{3}
(\tilde{p}_{03}-\tilde{p}_{23})(e_{33}-e_{22}).
\end{align*}
This way, we showed that $\sigma_{22}$ and $\sigma_{33}$ are not the eigenvectors of $\Lambda_{KF}(t)$, and hence $\Lambda_{KF}(t)$ is not the EV Gell-Mann channel.
\end{example}

\section{Time-local generators}

Analogously to the previous case, there are two natural ways to generalize the time-local generator of the Pauli channel (\ref{Pauli_generator}). The most well-known form of the time-local generator is the GKSL form. Let us consider the generator $\mathcal{L}_{LF}(t):\mathcal{B}(\mathcal{H}_n)\to\mathcal{B}(\mathcal{H}_n)$ with time-dependent decoherence rates $\gamma_{ij}(t)$ and the noise operators given by the Gell-Mann matrices,
\begin{equation}\label{generator_LF}
\mathcal{L}_{LF}(t)[X]=\sum_{i,j=0}^{n-1}\gamma_{ij}(t)
\left(\sigma_{ij}X\sigma_{ij}-\frac 12 \{\sigma_{ij}^2,X\}\right).
\end{equation}
The solution of (\ref{generator_LF}) is always a trace-preserving map. The condition for complete positivity is generally non-trivial and very hard to verify.

\begin{define}
The generator $\mathcal{L}_{LF}$ given by the GKSL form (\ref{generator_LF}) and whose decoherence rates $\gamma_{ij}(t)$ are chosen in such a way that it generates a CPT map is called the LF Gell-Mann generator.
\end{define}

Problems with generator $\mathcal{L}_{LF}(t)$ arise when we want to compare its solution to the Gell-Mann channels. Fortunately, there is another generator -- let us denote it by $\mathcal{L}_{EV}(t)$ -- for which this task is much easier. We define the generator by its eigenvalue equations,
\begin{equation}\label{generator_EV}
\mathcal{L}_{EV}(t)[\sigma_{ij}]=\eta_{ij}(t)\sigma_{ij}.
\end{equation}
The eigenvalues $\eta_{ij}(t)$ of $\mathcal{L}_{EV}(t)$ are in one-to-one correspondence with the eigenvalues $\lambda_{ij}(t)$ of $\Lambda_{EV}(t)$. The master equation (\ref{master_equation}) helps us to determine the one-to-one correspondence between $\eta_{kl} (t)$ and $\lambda_{kl} (t)$,
\begin{equation}\label{eta2lambda}
\eta_{kl} (t)=\frac{\der}{\der t}[\ln\lambda_{kl} (t)],
\end{equation}
as well as the inverse transformation,
\begin{equation}\label{lambda2eta}
\lambda_{kl} (t)=\exp\left[\int_0^t\eta_{kl} (\tau)\der \tau\right].
\end{equation}
Note that $\mathcal{L}_{EV}(t)$ describes a quantum evolution if and only if $\eta_{kl}(t)$ satisfy the conditions (\ref{CP_condition_1}, \ref{CP_condition_2}) with $\lambda_{kl}(t)$ given by (\ref{lambda2eta}).

\begin{define}
The generator $\mathcal{L}_{EV}$ given by the eigenvalue equations (\ref{generator_EV}) and for whose eigenvalues $\eta_{ij}(t)$ the conditions in (\ref{CP_condition_1}, \ref{CP_condition_2}, \ref{lambda2eta}) hold is called the EV Gell-Mann generator.
\end{define}

In the previous section, we showed that, under certain circumstances, the EV Gell-Mann channel has the corresponding Kraus form with the Gell-Mann matrices as its Kraus operators. Now, we would like to ask an analogical question for the generators: When do $\mathcal{L}_{LF}$ and $\mathcal{L}_{EV}$ generate the same dynamics? We answer this question in the following theorems.

\begin{thm}\label{gamma2eta}
The LF Gell-Mann generator $\mathcal{L}_{LF}$ is the EV Gell-Mann generator if and only if for its decoherence rates $\gamma_{ij}(t)$ the following $(n-1)(n-2)/2$ conditions hold:
\begin{equation}\label{gamma_cond}
\bigforall_{0\leq j,k< l\leq n-1}\quad \tilde{\gamma}_{jl}
=\tilde{\gamma}_{kl}=:\tilde{\gamma}_l,
\end{equation}
where $\tilde{\gamma}_{kl}:=\gamma_{kl}+\gamma_{lk}$ for $k<l$. Its eigenvalues can be found using the formulas listed below:
\begin{align}
&\eta_{kl}=-2\gamma_{lk}-\frac 12 \left[k\tilde{\gamma}_k+(l-1)\tilde{\gamma}_l
+\sum_{k<j<l}\tilde{\gamma}_j+2\sum_{l<j<n}\tilde{\gamma}_j\right]
-\sum_{k<j<l}\frac{1}{j(j+1)}\gamma_{jj}-\frac{k}{k+1}\gamma_{kk}
-\frac{l+1}{l}\gamma_{ll},\label{eta_SA}\\
&\eta_{kk}=-(k+1)\tilde{\gamma}_k-\sum_{k<j<n}\tilde{\gamma}_j\label{eta_D}.
\end{align}
\end{thm}

\begin{thm}\label{eta2gamma}
The EV Gell-Mann generator $\mathcal{L}_{EV}$ is the LF Gell-Mann generator if and only if $(n-1)(n-2)/2$ constraints listed below are satisfied:
\begin{align}
&\bigforall_{2\leq l\leq n-1}\quad \tilde{\eta}_{0l}=\tilde{\eta}_{1l},\label{eta_cond_1}\\
&\bigforall_{2\leq k\leq n-2}\quad
\eta_{kk}=\eta_{11}+
\sum_{0<j<k}\frac{j+1}{2}\left[-\frac{j+2}{j}(\tilde{\eta}_{j,j+2}-\tilde{\eta}_{j+1,j+2})
+\tilde{\eta}_{j-1,j}-\tilde{\eta}_{j-1,j+1}\right]\label{eta_cond_2},\\
&\bigforall_{1\leq k<l<m\leq n-1}\quad \tilde{\eta}_{kl}-\tilde{\eta}_{k-1,l}=\tilde{\eta}_{km}-\tilde{\eta}_{k-1,m},\label{eta_cond_3}
\end{align}
where $\tilde{\eta}_{kl}:=\eta_{kl}+\eta_{lk}$ for $k<l$. 
We can find the generator's decoherence rates from the following equations:
\begin{align}
\gamma_{kl}=&\frac 12 \left[-\frac{1}{l+1}\eta_{ll}
+\sum_{l<j<n}\frac{1}{j(j+1)}\eta_{jj}\right]
+\frac 14 (\eta_{kl}-\eta_{lk}),\label{gamma_SA}\\
\gamma_{kk}=&-\frac 14 \left[\tilde{\eta}_{k-1,k}-\sum_{k<j<n}
\frac{2}{j(j+1)}\eta_{jj}-\frac{2}{k(k+1)}\eta_{kk}\right]
+\frac 14 \frac{k-1}{k+1}
\left[\tilde{\eta}_{0,k-1}-\tilde{\eta}_{0k}+2\eta_{kk}\right]\nonumber
\\&+\frac{1}{2k}\frac{k-1}{k+1}\eta_{k-1,k-1}\label{gamma_D}.
\end{align}
\end{thm}

Again, it would be wise to provide a simple example of the Gell-Mann generator that is LF but not EV. This time, the lowest dimension where we can find such generators is $n=3$. Therefore, if the EV Gell-Mann channel is KF, it does not necessarily mean that its (EV) Gell-Mann generator is LF -- and vice versa. In short -- having the Kraus form (\ref{lambda_KF}) or the GKSL generator (\ref{generator_LF}) does not guarantee having the other.

\begin{example}
For $n=3$, the LF Gell-Mann generator is EV if and only if
\begin{equation}
\tilde{\gamma}_{02}=\tilde{\gamma}_{12}\qquad\mathrm{or}\qquad\gamma_{02}+\gamma_{20}=\gamma_{12}+\gamma_{21}.
\end{equation}
Now, let us take $\mathcal{L}_{LF}(t)$, which guarantees that the underlying quantum map is TP by definition, and arbitrary decoherence rates $\gamma_{ij}(t)$. Check the action of $\mathcal{L}_{LF}(t)$ on the Gell-Mann matrices:
\begin{align*}
&\mathcal{L}_{LF}[\sigma_{00}]=0,\\
&\mathcal{L}_{LF}[\sigma_{01}]=-\frac 12 (4\gamma_{10}+4\gamma_{11}+\tilde{\gamma}_{02}+\tilde{\gamma}_{12})\sigma_{01},\\
&\mathcal{L}_{LF}[\sigma_{10}]=-\frac 12 (4\gamma_{01}+4\gamma_{11}+\tilde{\gamma}_{02}+\tilde{\gamma}_{12})\sigma_{10},\\
&\mathcal{L}_{LF}[\sigma_{02}]=-\frac 12 (\tilde{\gamma}_{01}+\gamma_{11}+4\gamma_{20}+\tilde{\gamma}_{12}+3\gamma_{22})\sigma_{02},\\
&\mathcal{L}_{LF}[\sigma_{20}]=-\frac 12 (\tilde{\gamma}_{01}+\gamma_{11}+4\gamma_{02}+\tilde{\gamma}_{12}+3\gamma_{22})\sigma_{20},\\
&\mathcal{L}_{LF}[\sigma_{12}]=-\frac 12 (\tilde{\gamma}_{01}+\gamma_{11}+\tilde{\gamma}_{02}+4\gamma_{21}+3\gamma_{22})\sigma_{12},\\
&\mathcal{L}_{LF}[\sigma_{21}]=-\frac 12 (\tilde{\gamma}_{01}+\gamma_{11}+\tilde{\gamma}_{02}+4\gamma_{12}+3\gamma_{22})\sigma_{21},\\
&\mathcal{L}_{LF}[\sigma_{11}]=-\frac 12
\left[(4\tilde{\gamma}_{01}+\tilde{\gamma}_{02}+\tilde{\gamma}_{12})\sigma_{11}
+\sqrt{3}(\tilde{\gamma}_{02}-\tilde{\gamma}_{12})\sigma_{22}\right],\\
&\mathcal{L}_{LF}[\sigma_{22}]=-\frac 12
\left[\sqrt{3}(\tilde{\gamma}_{02}-\tilde{\gamma}_{12})\sigma_{11}
+3(\tilde{\gamma}_{02}+\tilde{\gamma}_{12})\sigma_{22}\right].
\end{align*}
Clearly, $\sigma_{11}$ and $\sigma_{22}$ are not the eigenvectors of $\mathcal{L}_{LF}(t)$, and therefore $\mathcal{L}_{LF}(t)$ is not the EV Gell-Mann generator.
\end{example}

\section{Applications and final thoughts}

In summary, we compared two possible definitions of the Gell-Mann channel: with the eigenvalue equations and the Kraus form. We derived the conditions for complete positivity and preserving the trace for these channels. We also showed when both channels describe the same dynamics and how to switch between the two forms. For the Gell-Mann channel given by its eigenvalue equations, we found its time-local generator together with the conditions under which this generator has the corresponding GKSL form.

The EV Gell-Mann channels can be used in the theory of quantum coherences and quantum correlation measures \cite{HuFan1, HuFan2}. It was shown that, under certain circumstances, the evolution equations of coherences and correlation measures obey the factorization relation. One of the conditions states that the eigenvectors of the quantum channel have to form an orthonormal basis. The three-dimensional Gell-Mann channel is listed as one of the examples, but it is obvious that the same theorems will hold for higher-dimensional EV Gell-Mann channels.

There are still some open questions which we failed to answer. The full comparative analysis of the EV and KF Gell-Mann channels won't be possible without deriving the eigenvalues of the general KF channel and LF generator. It would be interesting to check whether or not they describe the same dynamics, and -- if they don't, in general -- what are the conditions under which they do.

\section{Appendices}

\subsection{Trace-preserving map - Proof of Theorem \ref{TP_theorem}}

First, let us calculate the trace of $\Lambda_{KF}[X]$,
\begin{equation}
\Lambda_{KF}[X]=\sum_{0<j<n}\sum_{i<j}\tilde{p}_{ij}\Tr[X(e_{ii}+e_{jj})]
+\sum_{0<j<n}\frac{2}{j(j+1)}p_{jj}\Tr\left[X\left(\sum_{i<j}e_{ii}+j^2e_{jj}\right)\right]
+p_{00}\Tr[X],
\end{equation}
with $\tilde{p}_{ij}:=p_{ij}+p_{ji}$ for $i<j$. We want to have $\Tr\Lambda_{KF}[X]=\Tr[X]$ for all $X$, from which it follows that
\begin{equation}
\sum_{0<j<n}\sum_{i<j}\tilde{p}_{ij}(e_{ii}+e_{jj})
+\sum_{0<j<n}\frac{2}{j(j+1)}p_{jj}\left(\sum_{i<j}e_{ii}+j^2e_{jj}\right)
+(p_{00}-1)\sum_{0\leq j<n}e_{jj}=0.
\end{equation}
If we shuffle the indices and use the fact that $e_{ij}$ form an orthonormal basis, we arrive at
\begin{equation}\label{TPC}
p_{00}-1+\sum_{0\leq i<j}\tilde{p}_{ij}+\frac{2j}{j+1}p_{jj}+\sum_{j<i<n}
\left(\tilde{p}_{ji}+\frac{2}{i(i+1)}p_{ii}\right)=0.
\end{equation}
For $j=0$, equation (\ref{TPC}) simplifies to (\ref{TP_conditions1}),
\begin{equation*}
p_{00} =1-\sum_{0<l<n}\tilde{p}_{0l}-2\sum_{0<l<n}\frac{1}{l(l+1)}p_{ll}.
\end{equation*}
For $j=1$, it gives (\ref{TP_conditions2}),
\begin{equation*}
\sum_{1<l<n}(\tilde{p}_{1l}-\tilde{p}_{0l})=0.
\end{equation*}
For $j=2$, one arrives at (\ref{TP_conditions3}),
\begin{equation*}
p_{22}=p_{11}+\tilde{p}_{01}-\tilde{p}_{12}+\sum_{2<l<n}(\tilde{p}_{0l}-\tilde{p}_{2l}).
\end{equation*}
And, finally, for $j\geq 2$, we get
\begin{equation}
p_{jj}=\frac{j}{2(j-1)}\left[2\sum_{0<l<j}\frac{1}{l(l+1)}p_{ll}-\sum_{l<j}\tilde{p}_{lj} -\sum_{j<l<n}\tilde{p}_{jl}+\sum_{0<l<n}\tilde{p}_{0l}\right]\quad\bigforall_{1<j<n}.
\end{equation}
If we subtract the equation for $p_{jj}$ from the equality for $p_{j+1,j+1}$, then we obtain the recurrence relation,
\begin{equation}
p_{j+1,j+1}=p_{jj}-\frac{j+1}{2j}\left[
\sum_{0\leq l<j}(\tilde{p}_{l,j+1}-\tilde{p}_{lj})
+\sum_{j+1<l<n}(\tilde{p}_{j+1,l}-\tilde{p}_{jl})
\right]\quad\bigforall_{1<j<n}.
\end{equation}
Now, summing from $2$ to $k$ results in
\begin{equation}
\begin{split}
\sum_{j=2}^k(p_{j+1,j+1}-p_{jj})&=p_{k+1,k+1}-p_{22}\\&=-\sum_{j=2}^k\frac{j+1}{2j}
\left[\sum_{0\leq l<j}(\tilde{p}_{l,j+1}-\tilde{p}_{lj})
+\sum_{j+1<l<n}(\tilde{p}_{j+1,l}-\tilde{p}_{jl})
\right]\quad\bigforall_{2<k<n},
\end{split}
\end{equation}
which leads to (\ref{TP_conditions4}),
\begin{equation*}
p_{kk}=p_{22}+\sum_{1<j<k}\frac{j+1}{2j}\left[
\sum_{0\leq l<j}(\tilde{p}_{lj}-\tilde{p}_{l,j+1})
+\sum_{j+1<l<n}(\tilde{p}_{jl}-\tilde{p}_{j+1,l})
\right]\quad\bigforall_{2<k<n}.
\end{equation*}

\subsection{Completely positive map - Proof of Theorem \ref{CP_theorem}}

From Choi's theorem \cite{Choi}, it follows that a Hermitian map $\Lambda(t)$ with real eigenvalues $\lambda_\alpha(t)$ and eigenvectors $v_\alpha$ is completely positive if and only if \cite{Lee}
\begin{equation}
\sum_\alpha\lambda_\alpha\bar{v}_\alpha\otimes v_\alpha\geq 0,
\end{equation}
where $\bar{v}_\alpha$ denotes the complex conjugation of $v_\alpha$. In our case, the above condition reads
\begin{equation}\label{CP_cond}
\sum_{i,j=0}^{n-1}\lambda_{ij}(t)\bar{\sigma}_{ij}\otimes\sigma_{ij}\geq 0,
\end{equation}
which can be written as
\begin{equation}
\begin{split}
&\sum_{0<j<n}\sum_{0\leq i<j}\Bigg[
(\lambda_{ij}+\lambda_{ji})(e_{ij}\otimes e_{ij}+e_{ji}\otimes e_{ji})+
(\lambda_{ij}-\lambda_{ji})(e_{ij}\otimes e_{ji}+e_{ji}\otimes e_{ij})\\&+
\left(\lambda_{00}-\frac{2}{j+1}\lambda_{jj}+\sum_{j<k<n}\frac{2}{k(k+1)}
\lambda_{kk}\right)(e_{ii}\otimes e_{jj}+e_{jj}\otimes e_{ii})
+\frac{2}{j(j+1)}\lambda_{jj}e_{ii}\otimes e_{ii}
\Bigg]\\&
+\sum_{0\leq j<n}\left(\lambda_{00}+\frac{j}{j+1}\lambda_{jj}\right)
e_{jj}\otimes e_{jj}\geq 0
\end{split}
\end{equation}
after applying the definitions of the Gell-Mann matrices (\ref{Gell-Manns}). Let us introduce a new basis by $f_{in+k,jn+l}:=e_{ij}\otimes e_{kl}$. The matrix in (\ref{CP_cond}) splits into blocks formed by the basis vectors $f_{kl}$ sharing the $k$'th or $l$'th position. Our problem simplifies to checking the positivity of these blocks. The first block $A$ is a matrix $n\times n$ in the basis $\{f_{j(n+1),j(n+1)}\}_{0\leq j<n}$, constructed from
\begin{equation}\label{matrix_A}
\sum_{0\leq j<n}\left(\lambda_{00}+\frac{j}{j+1}\lambda_{jj}\right)
f_{j(n+1),j(n+1)}+\sum_{0<j<n}\sum_{0\leq i<j}\left(\lambda_{ij}+\lambda_{ji}
+\frac{2}{j(j+1)}\lambda_{jj}\right)f_{i(n+1),i(n+1)}.
\end{equation}
The other $n(n-1)/2$ blocks $B_{ij}$, $0\leq i<j<n$, are $2\times 2$ matrices in the respective bases $\{f_{in+j,in+j},f_{in+j,jn+i}, f_{jn+i,in+j},f_{jn+i,jn+i}\}$, and they can be constructed from
\begin{equation}\label{matrix_B}
(\lambda_{ij}+\lambda_{ji})(f_{in+j,jn+i}+f_{jn+i,in+j})+\left(\lambda_{00}
-\frac{2}{j+1}\lambda_{jj}+\sum_{j<k<n}\frac{2}{k(k+1)}\lambda_{kk}\right)
(f_{in+j,in+j}+f_{jn+i,jn+i}).
\end{equation}
Using equations (\ref{matrix_A}, \ref{matrix_B}), it is easy to find the matrix elements $A_{ij}$ of $A$ (\ref{A_ij}, \ref{A_jj}) and the form of $B_{ij}$,
\begin{equation*}
A_{ij}=\lambda_{ij}+\lambda_{ji},\qquad 
A_{jj}=\lambda_{00}+\frac{j}{j+1}\lambda_{jj}+\sum_{j<k<n}\frac{2}{k(k+1)}\lambda_{kk},
\end{equation*}
\begin{equation*}
B_{ij}=\begin{bmatrix}
\lambda_{00}-\frac{2}{j+1}\lambda_{jj}+\sum_{j<k<n}\frac{2}{k(k+1)}\lambda_{kk} & \lambda_{ij}-\lambda_{ji} \\
\lambda_{ij}-\lambda_{ji} & 
\lambda_{00}-\frac{2}{j+1}\lambda_{jj}+\sum_{j<k<n}\frac{2}{k(k+1)}\lambda_{kk}
\end{bmatrix}.
\end{equation*}
The condition $B_{ij}\geq 0$ implies (\ref{CP_condition_1}),
\begin{equation*}
|\lambda_{ij}-\lambda_{ji}|\leq|\lambda_{00}^D-\frac{2}{j+1}\lambda_{jj}
+\sum_{j<k<n}\frac{2}{k(k+1)}\lambda_{kk}|.
\end{equation*}

\subsection{KF channel that is EV - Proof of Theorem \ref{p2lambda}}

We start with the KF mapping; that is, we are given an arbitrary set of coefficients $p_{ij}(t)$. Through simple calculations, we show that only the Gell-Mann matrices $\sigma_{kl}$ with $k\neq l$ are the eigenvectors of $\Lambda_{KF}(t)$. The additional requirement for $\sigma_{kk}$ to be the eigenvectors of $\Lambda_{KF}(t)$ results in the following set of equations:
\begin{align}
-k\lambda_{kk} =&\sum_{0\leq j<k}\tilde{p}_{jk}-2k\left(\sum_{k<j<n}\frac{1}{j(j+1)}p_{jj}
+\frac{k}{k+1}p_{kk}\right),\label{first_ch}\\
\lambda_{kk} =&\sum_{l<j<n}\frac{2}{j(j+1)}p_{jj}+\frac{2l}{l+1}p_{ll} +\sum_{l\neq j<k}\tilde{p}_{jl}-k\tilde{p}_{kl}\quad\bigforall_{l<k},\label{second_ch}\\
k\tilde{p}_{kl}=&\sum_{0\leq j<k}\tilde{p}_{jl}\quad\bigforall_{l>k}\label{third_ch}.
\end{align}
For various values of $k$, equation (\ref{third_ch}) leads to
\begin{align*}
&k=1:\qquad\gamma_{1i}=\gamma_{0i},\\
&k=2:\qquad 2\gamma_{2i}=\gamma_{0i}+\gamma_{1i},\\
&k=3:\qquad 3\gamma_{3i}=\gamma_{0i}+\gamma_{1i}+\gamma_{2i},
\end{align*}
and so on, which implies (\ref{p_cond}),
\begin{align*}
\bigforall_{0\leq j,k<l\leq n-1}\quad
\tilde{p}_{jl}=\tilde{p}_{kl}=:\tilde{p}_l,
\end{align*}

Now, $\Lambda_{KF}(t)$ is a quantum channel if its coefficients $p_{ij}(t)$ satisfy
\[p_{ij}(t)\geq 0,\qquad p_{00}(0)=1,\qquad p_{ij}(0)=0,\]
together with the TP conditions (\ref{TP_conditions1}-\ref{TP_conditions4}). Note that the requirements for $p_{ij}(t)$ (\ref{p_cond}) guarantee that the second TP condition (\ref{TP_conditions2}) holds. Also, using (\ref{p_cond}) allows us to rewrite the rest of the TP conditions (\ref{TP_conditions2}-\ref{TP_conditions4}) in the following form,
\begin{align}
&p_{kk}=p_{11}+\frac 12 \left(\tilde{p}_1+\sum_{0<j<k}\tilde{p}_j-k\tilde{p}_k\right)\label{p_cond_2},\\
&p_{00}=1-\frac{n-1}{n}\left(p_{11}+\tilde{p}_1+\sum_{0<j<n}\tilde{p}_j\right)\label{p_cond_3}.
\end{align}
Straighforward calculations lead to $\lambda_{00}=1$ and $\lambda_{ij}(t)$ given in (\ref{lambda_SA}-\ref{lambda_D}).

\subsection{EV channel that is KF - Proof of Theorem \ref{lambda2p}}

The equation for $\lambda_{ij}$ (\ref{lambda_SA}) allows us to find
\begin{equation}\label{p_11^D}
\lambda_{kl}+\lambda_{lk}=2\left(1-2p_{11}-\tilde{p}_1-\sum_{j\neq l}\tilde{p}_j\right),
\end{equation}
from which follows the first condition for the eigenvalues (\ref{lambda_cond_1}),
\begin{equation*}
\bigforall_{0\leq j,k< l\leq n-1}\quad
\tilde{\lambda}_{jl}=\tilde{\lambda}_{kl}=:\tilde{\lambda}_l.
\end{equation*}

Now, let us take a look at the formula for $\lambda_{kk}$ (\ref{lambda_D});
\begin{align*}
&k=n-1:\qquad\lambda_{n-1,n-1}-1=-n\tilde{p}_{n-1}
\quad\implies\quad
\tilde{p}_{n-1}=-\frac 1n (\lambda_{n-1,n-1}-1),
\\[5mm]
&k=n-2:\qquad\lambda_{n-2,n-2}-1=-(n-1)\tilde{p}_{n-2}-\tilde{p}_{n-1}
\\&\implies\quad
\tilde{p}_{n-2}=-\frac{1}{n-1}(\lambda_{n-2,n-2}-1)+\frac{1}{n(n-1)}(\lambda_{n-1,n-1}-1),
\\[5mm]
&k=n-3:\qquad\lambda_{n-3,n-3}-1=-(n-2)\tilde{p}_{n-3}-\tilde{p}_{n-2}-\tilde{p}_{n-1}
\\&\implies\quad
\tilde{p}_{n-3}=-\frac{1}{n-2}(\lambda_{n-3,n-3}-1)+\frac{1}{(n-1)(n-2)}(\lambda_{n-2,n-2}-1)
+\frac{1}{n(n-1)}(\lambda_{n-1,n-1}-1).
\end{align*}
Hence, one obtains the equation for $\tilde{p}_k$,
\begin{equation}\label{p_k}
\tilde{p}_k=\frac 1n - \frac{1}{k+1}\lambda_{kk}
+\sum_{k<j<n}\frac{1}{j(j+1)}\lambda_{jj}.
\end{equation}
Using (\ref{p_k}) and the eigenvalue condition (\ref{lambda_cond_1}) in equation (\ref{p_11^D}), we rewrite $p_{11}$ in the following form:
\begin{equation}
p_{11}=\frac{1}{2n}-\frac 14 \tilde{\lambda}_l-\frac 12 \left(
\frac{1}{l+1}\lambda_{ll}-\lambda_{11}-\sum_{l<j<n}\frac{1}{j(j+1)}
\lambda_{jj}\right)\quad\bigforall_{0<l<n}.
\end{equation}
Comparing the formulas for $p_{11}$ for $l$ and $l-1$, we get the second eigenvalue condition (\ref{lambda_cond_2}),
\begin{equation*}
\bigforall_{1\leq k\leq n-1}\quad
\lambda_{kk} =\lambda_{11} +\frac 12 \left(\tilde{\lambda}_1+\sum_{0<j<k}\tilde{\lambda}_j-k\tilde{\lambda}_k\right).
\end{equation*}

Equation for $\tilde{p}_k$ (\ref{p_k}), along with
\begin{equation}
p_{kl}-p_{lk}=\frac 12 \left(\lambda_{kl}-\lambda_{lk}\right)
\end{equation}
derived from $\lambda_{kl}$ (\ref{lambda_SA}), gives us the formula for $p_{kl}$ (\ref{p_SA}). We obtain $p_{kk}$ and $p_{00}$ (\ref{p_D}, \ref{p_0}) from (\ref{p_cond_2}, \ref{p_cond_3}).

\subsection{LF generator that is EV - Proof of Theorem \ref{gamma2eta}}

Let us start with the LF Gell-Mann generator $\mathcal{L}_{LF}(t)$. Simple calculations show that $\sigma_{kl}$ are the eigenvectors of $\mathcal{L}_{LF}(t)$, whereas $\sigma_{kk}$ are not. The requirements for $\sigma_{kk}$ to be the eigenvectors of the LF Gell-Mann generator lead to the following set of equations:
\begin{align}
&-k\eta_{kk}=(k+1)\sum_{0\leq j<k}\tilde{\gamma}_{jk}
+k\sum_{k<j<n}\tilde{\gamma}_{kj},\label{first}\\
&\eta_{kk}=-(k+1)\tilde{\gamma}_{lk}-\sum_{k<j<n}\tilde{\gamma}_{lj}\quad\bigforall_{l<k},\label{second}\\
&k\tilde{\gamma}_{ki}=\sum_{0\leq j<k}\tilde{\gamma}_{ji}\quad
\bigforall_{i>k}\label{third}.
\end{align}
For various values of $k$, equation (\ref{third}) simplifies to
\begin{align*}
&k=1:\qquad\tilde{\gamma}_{1i}=\tilde{\gamma}_{0i},\\
&k=2:\qquad 2\tilde{\gamma}_{2i}=\tilde{\gamma}_{0i}+\tilde{\gamma}_{1i},\\
&k=3:\qquad 3\tilde{\gamma}_{3i}
=\tilde{\gamma}_{0i}+\tilde{\gamma}_{1i}+\tilde{\gamma}_{2i},
\end{align*}
which leads to (\ref{gamma_cond}),
\begin{equation*}
\bigforall_{0\leq j,k< l\leq n-1}\quad \tilde{\gamma}_{jl}
=\tilde{\gamma}_{kl}=:\tilde{\gamma}_l.
\end{equation*}
Equations (\ref{first}) and (\ref{second}) can be combined together to give
\begin{equation}
(k+1)(\tilde{\gamma}_{jk}-\tilde{\gamma}_{lk})=-\sum_{k<i<n}(\tilde{\gamma}_{ji}-\tilde{\gamma}_{li}),
\end{equation}
which, after applying condition (\ref{gamma_cond}), is always satisfied and yields to no additional constraints. Now, we can easily obtain the formulas for $\eta_{ij}$ (\ref{eta_SA}) and $\eta_{jj}$ (\ref{eta_D}).

\subsection{EV generator that is LF - Proof of Theorem \ref{eta2gamma}}

First, we would like to analyze the formula for $\eta_{kk}$ (\ref{eta_D});
\begin{align*}
&k=n-1:\qquad\eta_{n-1,n-1}=-n\tilde{\gamma}_{n-1}
\quad\implies\quad
\tilde{\gamma}_{n-1}=-\frac 1n \eta_{n-1,n-1},
\\[0.5cm]
&k=n-2:\qquad\eta_{n-2,n-2}=-(n-1)\tilde{\gamma}_{n-2}-\tilde{\gamma}_{n-1}
\\&\implies\quad
\tilde{\gamma}_{n-2}=-\frac{1}{n-1}\eta_{n-2,n-2}+\frac{1}{n(n-1)}\eta_{n-1,n-1},
\\[0.5cm]
&k=n-3:\qquad\eta_{n-3,n-3}=-(n-2)\tilde{\gamma}_{n-3}-\tilde{\gamma}_{n-2}-\tilde{\gamma}_{n-1}
\\&\implies\quad
\tilde{\gamma}_{n-3}=-\frac{1}{n-2}\eta_{n-3,n-3}+\frac{1}{(n-1)(n-2)}\eta_{n-2,n-2}
+\frac{1}{n(n-1)}\eta_{n-1,n-1}.
\end{align*}
From the above, we conclude that
\begin{equation}\label{gamma_l}
\tilde{\gamma}_l=-\frac{1}{l+1}\eta_{ll}+\sum_{l<j<n}\frac{1}{j(j+1)}\eta_{jj}.
\end{equation}
Moreover, the equation for $\eta_{ij}$(\ref{eta_SA}) implies
\begin{equation}\label{gamma_kl}
\gamma_{kl}-\gamma_{lk}=\frac 12 \left(\eta_{kl}-\eta_{lk}\right).
\end{equation}
By adding and subtracting (\ref{gamma_l}) and (\ref{gamma_kl}), we get the formula for $\gamma_{kl}$ (\ref{gamma_SA}).

Now, we are going to further analyze $\eta_{kl}$ (\ref{eta_SA}). Observe that for $k<l$
\begin{equation}
\begin{split}
\tilde{\eta}_{kl}:&=\eta_{kl}+\eta_{lk}\\&=-\left[
k\tilde{\gamma}_k+(l+1)\tilde{\gamma}_l+\sum_{k<j<l}\tilde{\gamma}_j+2\sum_{j>l}\tilde{\gamma}_j
+\sum_{k<j<l}\frac{2}{j(j+1)}\gamma_{jj}+\frac{2k}{k+1}\gamma_{kk}
+\frac{2(l+1)}{l}\gamma_{ll}
\right].
\end{split}
\end{equation}
Introduce a new symbol
\begin{equation}\label{alpha}
\begin{split}
\alpha_{kl}:&=-2\left[
\sum_{k<j<l}\frac{1}{j(j+1)}\gamma_{jj}+\frac{k}{k+1}\gamma_{kk}
+\frac{l+1}{l}\gamma_{ll}
\right]\\&
=\tilde{\eta}_{kl}+k\tilde{\gamma}_k+(l+1)\tilde{\gamma}_l+\sum_{k<j<l}\tilde{\gamma}_j+2\sum_{l<j<n}\tilde{\gamma}_j.
\end{split}
\end{equation}
From definition, we calculate
\begin{equation}
\alpha_{0,k+1}=-2\left[
\sum_{0<j<k}\frac{1}{j(j+1)}\gamma_{jj}+\frac{1}{k(k+1)}\gamma_{kk}
+\frac{k+2}{k+1}\gamma_{k+1,k+1}
\right],
\end{equation}
\begin{equation}
\alpha_{k,k+1}=-2\left[
\frac{k}{k+1}\gamma_{kk}+\frac{k+2}{k+1}\gamma_{k+1,k+1}
\right],
\end{equation}
\begin{equation}
\alpha_{0k}=-2\left[
\frac{k+1}{k}\gamma_{kk}+\sum_{0<j<k}\frac{1}{j(j+1)}\gamma_{jj}
\right].
\end{equation}
We can express $\gamma_{kk}$ in terms of $\alpha_{kl}$,
\begin{equation}
\gamma_{kk}=-\frac 14 \left(\alpha_{k,k+1}+\alpha_{0k}-\alpha_{0,k+1}\right).
\end{equation}
Using the r.h.s. of (\ref{alpha}) together with (\ref{gamma_l}), we find the formula for $\gamma_{jj}$ (\ref{gamma_D}).

To get the constraints for the eigenvalues $\eta_{kl}$ (\ref{eta_cond_1}, \ref{eta_cond_2}), we express $\alpha_{kl}$ in terms of $\tilde{\eta}_{kl}$,
\begin{equation}\label{alpha2}
\alpha_{kl}=\tilde{\eta}_{kl}-\frac{k}{k+1}\eta_{kk}-\frac{l^2+1}{l(l+1)}\eta_{ll}
-2\sum_{l<j<n}\frac{1}{j(j+1)}\eta_{jj}-\sum_{k<j<l}\frac{1}{j(j+1)}\eta_{jj}.
\end{equation}
Different values of $k$ lead to
\begin{align}
&k=0:\\&\alpha_{0l}:=
-2\frac{l+1}{l}\gamma_{ll}-2\sum_{0<j<l}\frac{1}{j(j+1)}\gamma_{jj}\nonumber\\&\quad\quad
=\tilde{\eta}_{0l}-\frac{l^2+1}{l(l+1)}\eta_{ll}-2\sum_{l<j<n}\frac{1}{j(j+1)}\eta_{jj}
-\sum_{0<j<l}\frac{1}{j(j+1)}\eta_{jj},\label{0l}
\\
&k=1:\\&\alpha_{1l}:=
-\gamma_{11}-2\frac{l+1}{l}\gamma_{ll}-2\sum_{1<j<l}\frac{1}{j(j+1)}\gamma_{jj}\nonumber\\&\quad\quad
=\tilde{\eta}_{1l}-\frac 12 \eta_{11}-\frac{l^2+1}{l(l+1)}\eta_{ll}
-2\sum_{l<j<n}\frac{1}{j(j+1)}\eta_{jj}
-\sum_{1<j<l}\frac{1}{j(j+1)}\eta_{jj},\label{1l}
\\
&k=l-1:\\&\alpha_{l-1,l}:=
-2\frac{l-1}{l}\gamma_{l-1,l-1}-2\frac{l+1}{l}\gamma_{ll}\nonumber\\&\quad\quad
=\tilde{\eta}_{l-1,l}-\frac{l-1}{l}\eta_{l-1,l-1}-\frac{l^2+1}{l(l+1)}\eta_{ll}
-2\sum_{l<j<n}\frac{1}{j(j+1)}\eta_{jj},\label{l-1,l}
\\
&k=l-2:\\&\alpha_{l-2,l}:=
-2\frac{l-2}{l-1}\gamma_{l-2,l-2}-2\frac{l+1}{l}\gamma_{ll}-2\frac{1}{l(l-1)}\gamma_{l-1,l-1}\nonumber\\&\quad\quad
=\tilde{\eta}_{l-2,l}-\frac{l-2}{l-1}\eta_{l-2,l-2}-\frac{l^2+1}{l(l+1)}\eta_{ll}
-2\sum_{l<j<n}\frac{1}{j(j+1)}\eta_{jj}-\frac{1}{l(l-1)}\eta_{l-1,l-1},\label{l-2,l}
\end{align}
where -- in each line -- the first equality comes from the definition of $\alpha_{kl}$ (\ref{alpha}), and the second one follows from (\ref{alpha2}).
Subtracting the respective sides of equations (\ref{0l}) and (\ref{1l}), we get the first constraint for the eigenvalues (\ref{eta_cond_1}),
\begin{equation*}
\bigforall_{2\leq l\leq n-1}\quad \tilde{\eta}_{0l}=\tilde{\eta}_{1l}.
\end{equation*}

After subtracting the respective sides of equations (\ref{l-1,l}) and (\ref{l-2,l}), as well as rearranging the indices, we obtain
\begin{equation}\label{cond_eta_first_eq}
2(\gamma_{kk}-\gamma_{k-1,k-1})=\frac{k}{k-1}(\tilde{\eta}_{k-1,k+1}-\tilde{\eta}_{k,k+1})
+\eta_{kk}-\eta_{k-1,k-1}\qquad\bigforall_{1<k<n-1}.
\end{equation}
On the other hand, different values of $l$ give us
\begin{align}
&l=k+1:\\&\alpha_{k,k+1}:=
-2\frac{k}{k+1}\gamma_{kk}-2\frac{k+2}{k+1}\gamma_{k+1,k+1}\nonumber\\&\quad\qquad
=\tilde{\eta}_{k,k+1}-\frac{k}{k+1}\eta_{kk}-\frac{(k+1)^2+1}{(k+1)(k+2)}\eta_{k+1,k+1}
-2\sum_{k+1<j<n}\frac{1}{j(j+1)}\eta_{jj},\label{k,k+1}
\\
&l=k+2:\\&\alpha_{k,k+2}:=
-2\frac{k}{k+1}\gamma_{kk}-2\frac{k+3}{k+2}\gamma_{k+2,k+2}
-2\frac{1}{(k+1)(k+2)}\gamma_{k+1,k+1}\nonumber\\&\qquad\quad
=\tilde{\eta}_{k,k+2}-\frac{k}{k+1}\eta_{kk}-\frac{(k+2)^2+1}{(k+2)(k+3)}\eta_{k+2,k+2}\nonumber\\&\quad\qquad
-2\sum_{k+2<j<n}\frac{1}{j(j+1)}\eta_{jj}-\frac{1}{(k+1)(k+2)}\eta_{k+1,k+1}\label{k,k+2},
\end{align}
where again the first equality comes from the definition of $\alpha_{kl}$ (\ref{alpha}), and the second one follows from (\ref{alpha2}).
Subtracting the respective sides of equations (\ref{k,k+1}) and (\ref{k,k+2}), as well as rearranging the indices, results in
\begin{equation}\label{cond_eta_second_eq}
2(\gamma_{kk}-\gamma_{k-1,k-1})=\frac{k}{k+1}(\tilde{\eta}_{k-2,k-1}-\tilde{\eta}_{k-2,k})
+\frac{k-1}{k+1}(\eta_{kk}-\eta_{k-1,k-1})\qquad\bigforall_{1<k<n-1}.
\end{equation}
If we compare (\ref{cond_eta_first_eq}) with (\ref{cond_eta_second_eq}), we arrive at the second condition for the eigenvalues (\ref{eta_cond_2}),
\begin{equation*}
\bigforall_{2\leq k\leq n-2}\quad
\eta_{kk}=\eta_{11}+
\sum_{0<j<k}\frac{j+1}{2}\left[-\frac{j+2}{j}(\tilde{\eta}_{j,j+2}-\tilde{\eta}_{j+1,j+2})
+\tilde{\eta}_{j-1,j}-\tilde{\eta}_{j-1,j+1}\right].
\end{equation*}

The last requirement for $\eta_{kl}(t)$ (\ref{eta_cond_3}) can be obtained by calculating $\alpha_{kl}-\alpha_{k-1,l}$ using equation (\ref{alpha2}). This way, we get
\begin{equation}
\tilde{\eta}_{kl}-\tilde{\eta}_{k-1,l}=2\frac{k-1}{k}
\left[\frac 12 (\tilde{\eta}_{kk}-\tilde{\eta}_{k-1,k-1})
-\gamma_{kk}+\gamma_{k-1,k-1}
\right],
\end{equation}
where the r.h.s. depends only on the index $k$. Therefore, one arrives at (\ref{eta_cond_3}),
\begin{equation*}
\bigforall_{1\leq k<l<m\leq n-1}\quad \tilde{\eta}_{kl}-\tilde{\eta}_{k-1,l}=\tilde{\eta}_{km}-\tilde{\eta}_{k-1,m}.
\end{equation*}

\end{document}